\documentclass[12pt]{article} \textheight 23cm
\usepackage{graphicx}
\usepackage{amsmath}
\begin{document} \begin{center}
{\large \bf Desiccation of a clay film: Cracking versus peeling }\\ \vskip 0.5cm
Supti Sadhukhan$^1$, Janett Prehl$^2$, Peter Blaudeck$^2$, K. H. Hoffmann$^2$,  Tapati Dutta$^3$ 
and Sujata Tarafdar$^4$\\
\vskip 0.5cm

$^1$ Physics Department, Jogesh Chandra Chaudhuri College,
\\ Kolkata 700033, India\\
$^2$ Institut f\"ur Physik,
            Technische Universit\"at, 09107 Chemnitz, Germany  \\
$^3$ Physics Department, St. Xavier's College, \\Kolkata 700016, India\\
$^4$ Condensed Matter Physics Research Centre,\\Physics Department, Jadavpur University,\\ Kolkata 700032, India\\

\end{center}
\vskip 1cm
\noindent {\bf Abstract}\\
\noindent Cracking and peeling of  a layer of clay on desiccation has been 
simulated using a spring model. A vertical section through the layer with finite thickness
is represented by a rectangular array of nodes connected by linear springs on a square lattice.
The effect of reduction of the natural length of the springs, which mimics the drying is studied.
Varying the strength of adhesion between sample and substrate and the rate of penetration of the
 drying front produces an interesting phase diagram, showing cross-over from peeling to cracking
behavior. Changes in the number and width of cracks on varying the layer thickness is  observed
to reproduce experimental reports.\\
\noindent {\bf PACS Nos:} 61.43.Bn - molecular dynamics in structural modelling, 62.20.mt - structural
failure.
\vskip 1cm

Formation of crack patterns in clay films is a familiar sight and is now a widely studied subject
of research.
Laboratory experiments \cite{coffee,shor,born,pats,dm1,dm2,dm3}, computer simulation and analytical studies \cite{kit,col1,col2,sadhu} and also
observations in natural surroundings \cite{wein} have been reported in considerable detail.
The present work is a spring model simulation, attempting to understand 
a different aspect of fracture of a film. 

On drying, a thin layer of clay or a coating of paint may be observed to crack 
while remaining attached
to the substrate, or alternatively it may `peel' off the substrate first, instead of cracking. 
Which process dominates, depends on external factors such as the temperature and humidity, 
which determine the
drying rate, as well as the strength of adhesion to the substrate and thickness of the layer.
 The present simulation investigates this angle of the desiccation problem. A molecular dynamics 
approach has been used to calculate
the dynamics of the system under the forces due to Hookean springs arranged on a square lattice.

Our system is a rectangular array of points on a square lattice, of length $L$ and height $H$ 
representing 
a vertical section through
the layer of clay. The nodes are connected by Hookean springs, initially having unit natural
length. The lowest row of nodes represents the substrate. All vertical and horizontal springs
have the same spring constant assumed to be unity in arbitrary units. All the springs except the lowest vertical row
have a breaking threshold $S1$ and break when the strain  exceeds this value. The lowest row
of vertical springs, connecting the sample to the substrate has a different threshold $S2$
representing the adhesion between sample and substrate.

Desiccation is implemented through a reduction in the natural length of the springs, so that the
system becomes strained. Desiccation starts at the top layer, since moisture evaporates from the 
surface.
There are two time scales involved here - desiccation continues at the top layer at every
timestep, with a magnitude decreasing successively and the desiccation is transmitted to the lower
layer after $t_d$ timesteps. So, $t_d = 1$ represents the fastest drying rate, when the interface
between dry and saturated clay moves downward rapidly. $t_d$ can be increased to mimic a slower 
drying. 

The shrinking due to drying decreases successively according to the following rule
\begin{equation} d_n = \exp (br^{ -n}/\ln r) \end{equation} 
$d_n$ represents the natural length of the drying spring at the $n$th timestep. $d_n$ is normalized
to $d_0$ at $n$=0. $b$ has been assigned the value 0.05 and $r = 1.1$ in this paper. 
When $n$ is large enough there is hardly any change in the subsequent natural lengths which saturates to a mimimum value $d_{min}$. The 
parameters $b$ and $r$, are chosen so that finally $d_{min}$ saturates to a value about 60-70\% of $d_0$.
This is the amount of shrinking usually observed.

The molecular dynamics proceeds as follows. The $r$th `layer' of particles is defined as the
$r$th horizontal row of springs from the bottom, together with the adjacent $(r+1)$th vertical
row of springs. The topmost layer of course, consists only of horizontal springs. So the
sample dries layer by layer, starting at the top. After the natural length of one layer is
reduced, the system relaxes to its minimum energy configuration. 

The force on each particle due to its neighboring springs is calculated and the maximum force on a particle noted.The force gives the acceleration $a$ for a paticle. A simplified form of Verlet's algorithm is used, whence
the particle at $x_n$ moves to $x_{n+1}$ given by

\begin{equation} x_{n+1} = x_n + a{\delta t}^2 \end{equation}
The time interval $\delta t$ has to be chosen judiciously. This is the time after which the
system is updated. If it is too small, a large number of intervals $\delta t$ are required
to make a finite timestep and the error for each interval accumulates to a considerable 
value. On the other hand, if the interval is too large, obviously the assumption of constant
acceleration during this interval breaks down and results are not reliable. Here we find that
assigning $\delta t=0.005$ gives realistic results. After every $\delta t$, the maximum force on a particle is noted. If the maximum force for successive $\delta t's$ ~ $10^{-7}$, the system is assumed to have relaxed completely. We then check whether the strain on any spring has exceeded the threshold, in which case it breaks.
If a number of springs cross the threshold simultaneously, the one with the highest strain
breaks. If again, there are more than one springs with the same highest strain, the central 
one is chosen to break. This situation arises sometimes at the first breaking, but it rarely 
arises later.

The molecular dynamics runs again, at the next timestep with the implementation of equation (1). 
This goes on until $d_min$ is reached.

For the row of lowest  vertical springs, which attach the sample to the substrate there is
in addition to breaking, a provision for horizontal {\it slipping}. 
A simple harmonic oscillator potential is assigned to this row of nodes. This tends to prevent
motion of these nodes along the vertical axis. If the force along the horizontal direction exceeds
the slipping threshold $V_{th}$, then the relaxed position, within the allowed limit is determined by molecular dynamics.

The subsequent desiccation is implemented, on the top layer (i.e. the $H$th layer) for $t_d$ times
before the desiccation is allowed to penetrate to the next lower layer, (i.e. the $(H-1)$th 
layer). So when any given layer is undergoing the $(p+t)$th drying step, the layer below it is
in the $p$th drying step. The whole process is continued until subsequent desiccations leave
the natural length of the springs almost unchanged. We now consider the sample to be completely
dry. The development of the sample is displayed graphically to note how its appearance changes
and how cracking and peeling proceed. The timestep when the crack or peel completes is also 
noted. A crack completes when it breaks the system into two disjoint pieces and the peel
completes when the sample is completely detached from the substrate. A crack may be vertical
or horizontal. When the sample detaches from the lowest row of nodes, we term the process as
{\it peeling}, whereas if it splits horizontally at any higher level we call it {\it horizontal
cracking}. 

Molecular dynamics has been run  with $L=20$ and $L=60$ for $H=8$ for a range of parameters
$S2$ varying from 0.02 to 0.5, with $S1=0.1$. However for $S2>0.5$, the crack pattern does not change. 
The time lag $t_d$ has been varied from 1 to 80.
A small set of samples with $L=80$ has been run to produce several stages of hierarchical
cracking with $H$= 4,8,12 to observe the effect of changing sample thickness. The slipping
threshold $V_{th}$ is either infinite (no slipping) or kept at 0.0001, to compare the results 
for a rough and a smooth substrate.

\begin{figure}[ht]
\begin{center}
\includegraphics[width=14.0cm, angle=270]{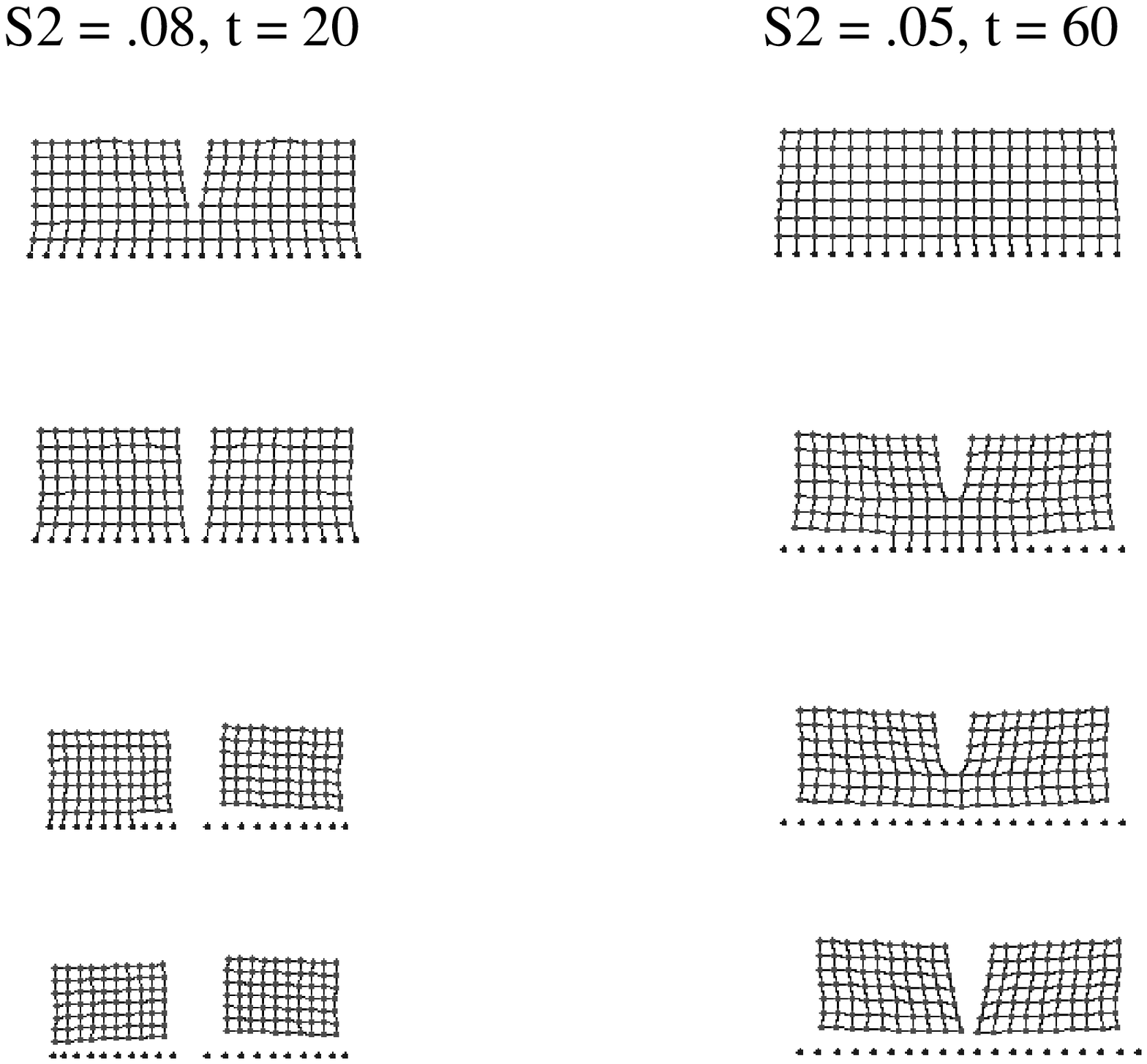}
\end{center}
\caption{Cracks formed in $L=20$ samples. Several successive stages are shown until desiccation
is complete. The left column shows cracking whereas the right column peels first, though cracking
initiated earlier.                                                                                
 } \label{short}
\end{figure}

\begin{figure}[ht]
\begin{center}
\includegraphics[width=14.0cm, angle=270]{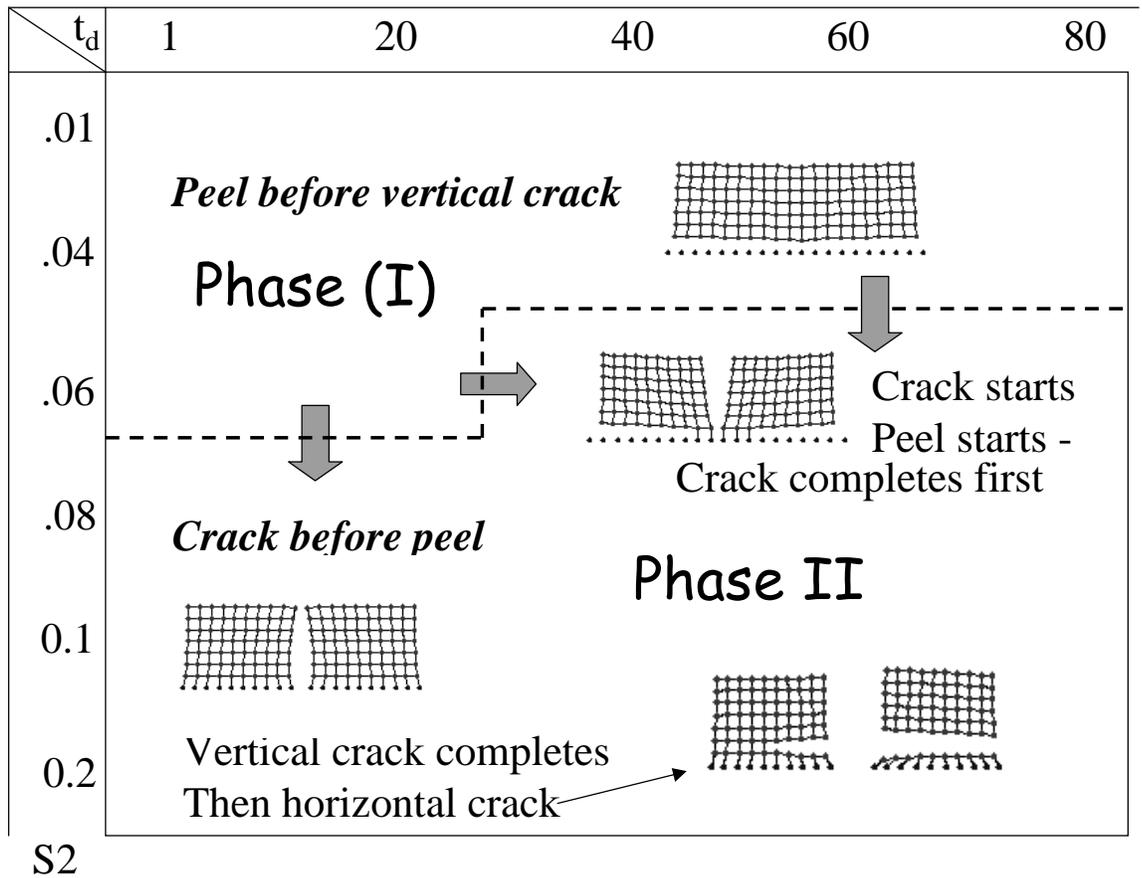}
\end{center}
\caption{ The phase diagram as $S2$ and $t_d$ vary are shown. Typical configurations for crack,
peel and simultaneous crack-peel are shown for system size L=20. The broken line separates Phase (I), where peeling
predominates and Phase (II), where vertical cracks bisect the system first. For increasing $S2$ and increasing $t_d$
the arrows indicate where  cross-over from peeling
to cracking occurs.                                                                                                     
 } \label{phased}
\end{figure}

\begin{figure}[ht]
\begin{center}
\includegraphics[width=14.0cm, angle=270]{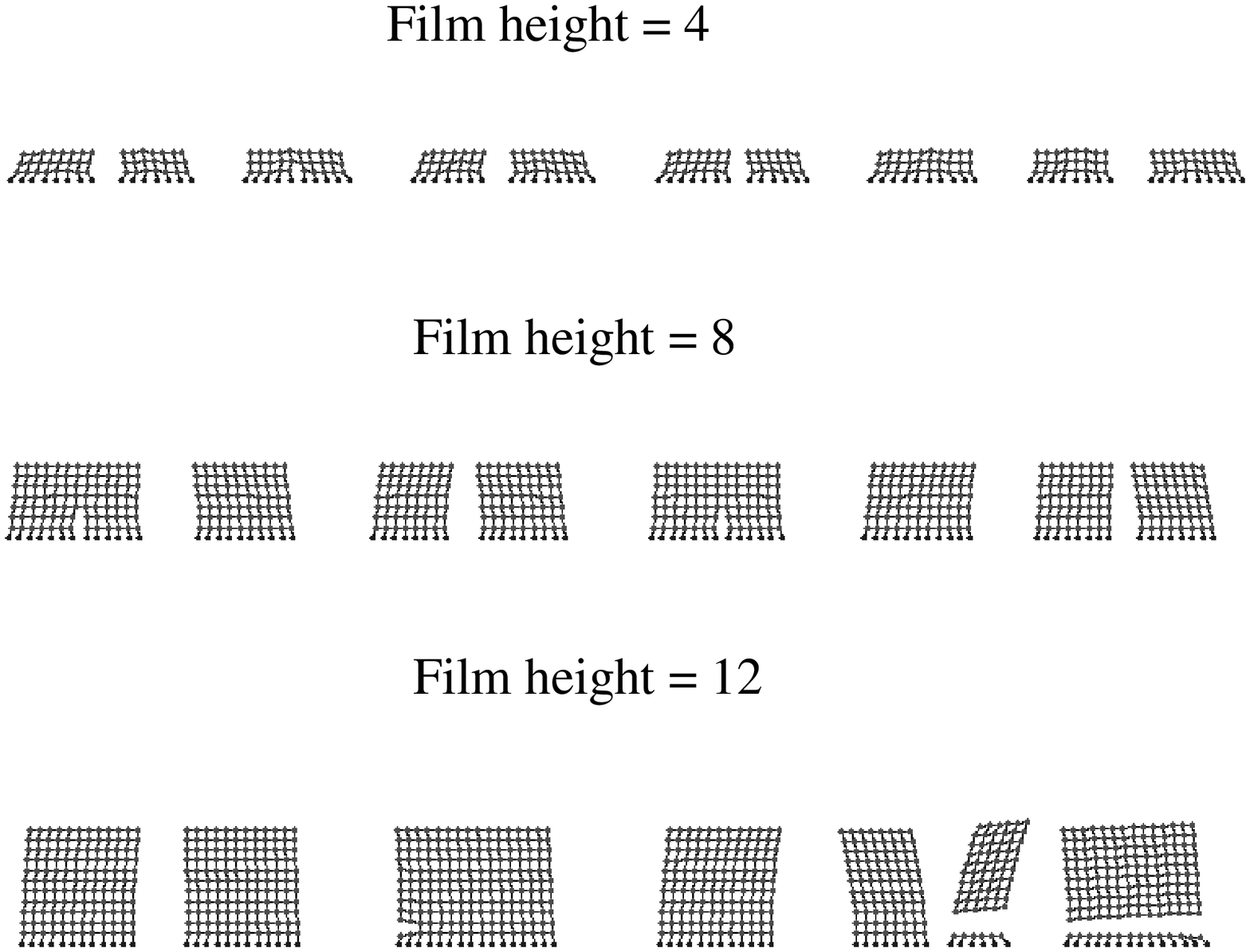}
\end{center}
\caption{ Final crack patterns in samples with $L=80$ and $H$ = 4, 8 and 12 from top to bottom.                                                                                                     
 } \label{h3}
\end{figure}

Figure \ref{short} shows several stages of crack development in $L=20$ samples. The results 
for  the full range of parameters are summarized in Figure \ref{phased}. The effect of 
varying $H$ for the long samples is illustrated in Figure \ref{h3}. 

In Figure \ref{short} we show two parameter sets, one of which cracks first and one which peels
first. Here the slipping threshold is low, this is clearly evident in the left column, where we see the
gap between the lowest nodes attached to the substrate has widened at the center. In the right
hand column, where $S2$ is smaller, peeling becomes easy because of low adhesion, so the sample
peels without slipping or cracking. 

The crack-peel variation over a wide range of parameters
is summarized in the phase diagram in Figure \ref{phased}. We find two transitions from 
peel to crack behavior, (i) as $S2$ increases at low $t_d$ and (ii) as $t_d$ increases for 
certain constant $S2$ values. The physics behind each is quite clear - in (i) as $S2$ increases, adhesion
to the substrate hinders peeling, making cracking more favourable. In (ii) large $t_d$ i.e. slow
drying, allows the system to relax during desiccation, suppressing distortion of the upper 
layers which would tend to curl up. This prevents peeling and strain has to be relieved by
cracking. Sometimes for  $S2>S1$, horizontal cracks also appear at large $t_d$ to accommodate
vertical strain, since peeling from the substrate is virtually prohibited here. The crack profile
is seen to taper from top to bottom or vice versa, depending on whether the crack starts at the
top or bottom.

The cross-over regions indicated in Figure \ref{phased}, change if the slip along the 
substrate is not allowed. We show the difference in cross-over regime for low and high slip 
threshold in table(\ref{slip}). There is also a change in the cross-over with $t_d$ for constant
$S2$ with sample size $L$. For $L=60$, the cross-over regime from peel-crack remained at 
$S2=0.06$ for all $t_d$. But no cross-over was observed on increasing $t_d$ at a constant $S2$. It 
would be iteresting to see if such a cross-over appears for $t_d>80$. 

\begin{figure}[ht]
\begin{center}
\includegraphics[width=14.0cm, angle=270]{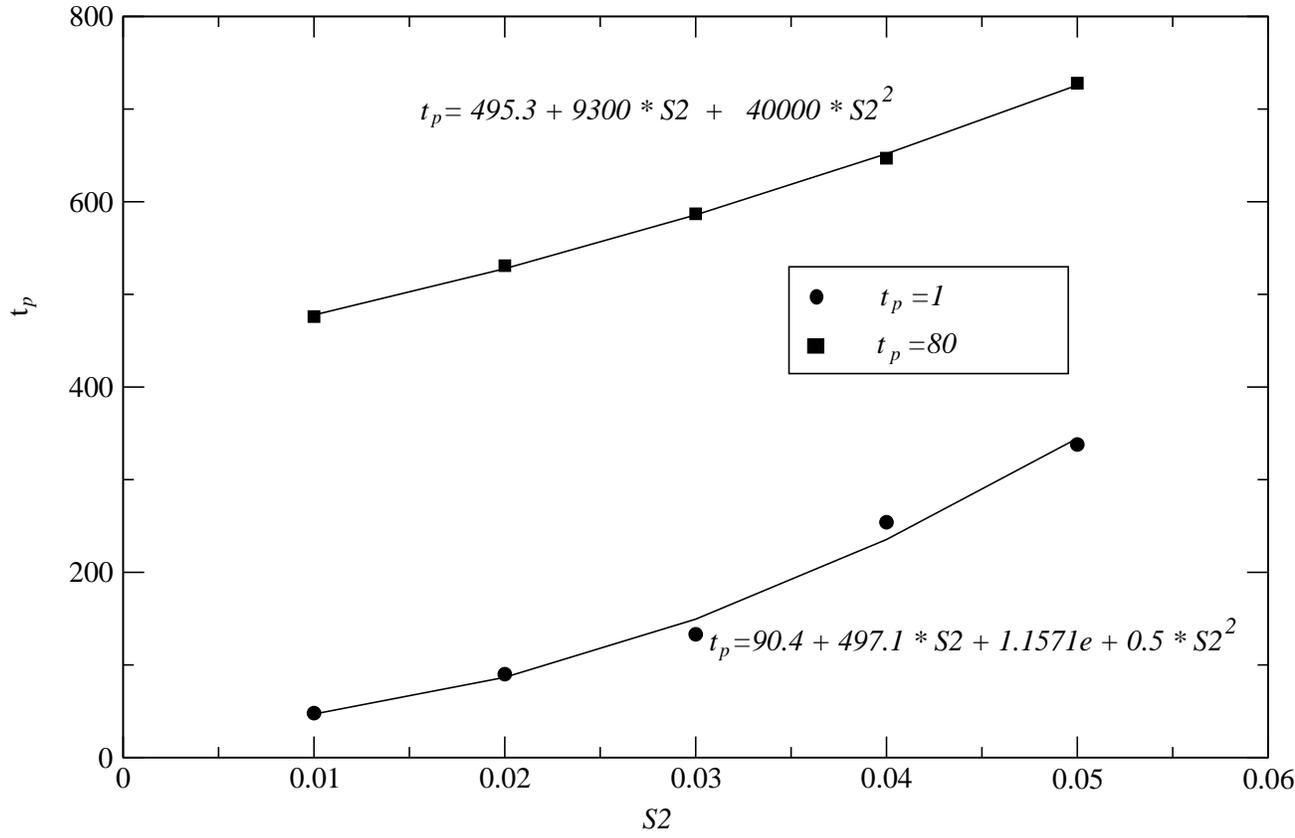}
\end{center}
\caption{ Variation in $t_p$, the peeling time as $S2$ increases from the lowest value and
approaches cross-over. The upper curve is for $t_d=80$ and the lower for $t_d=1$, the curve shows a
quadratic fit. All quantities are
in arbitrary units.                                                                                                      
 } \label{tp}
\end{figure}

As $S2$ increases for a given $t_d$ and the system approaches the transition from peel to crack,
we observe that the number of time steps required for peeling to complete goes on increasing upto
the cross-over. The variation in the time required for complete peeling $t_p$, is plotted in
Figure \ref{tp} for $t_d=1$  and $t_d=80$. It is seen that the data points show a good fit to a 
quadratic curve with the form
\begin{equation} t_p = A + BS_2 + C{S_2}^2 \end{equation}
Here A,B and C are constants.
It is natural that the time to complete  a process takes longer as a phase transition is 
approached, but we cannot explain the quadratic dependence at present. A further interesting
observation is that the time for a vertical crack to complete $t_c$, remains nearly constant
as we move upwards from $S2=0.2$ towards the  cross-over point.
We simulated desiccation of a series of $L=80$ samples of different thickness as well. The final
desiccated samples are shown in Figure \ref{h3}. A very well known feature of desiccation cracks
is well reproduced here, thicker samples have less number of wider cracks compared to thin samples.
For $H=4$, the number of cracks is 9, for $H=8$ it is 7 and for $H=12$ it is only 6. The widths 
of cracks also decrease with $H$, this has been reported in many experimental papers 
\cite{shor,dm3}.

Our results are realistic and this kind of behavior has been observed experimentally. 
In experiments on drying laponite
peeling was found to occur, with peds getting completely detached from the substrate, when the
desiccation was rapid due to low humidity (less than $\sim 50\%$) \cite{dm3}. Quantitative 
studies of the
effect of temperature and humidity under controlled conditions are yet to be done. Further, the parameters of the simulation
must be related to physical properties of sample and substrate, such as the viscosity of the 
clay suspension, surface roughness of the substrate and other factors affecting adhesion, e.g. 
the dielectric
constants of clay and substrate as well \cite{epje}. This will allow a precise comparison of
the simulation results with experiments. 
But even with the present simplistic approach our results are highly interesting and realistic. 

Earlier work \cite{sadhu} on Monte Carlo simulation
of a quasi-1-dimensional system studied the variation of area covered by the cracks observed under
different resolution of length scales for various substrates. This work covers a different aspect
in a 2-dimensional system. We hope to extend this work further to simulate the top view of the
crack pattern in future.

{\bf Acknowledgement} SS sincerely thanks the DFG for funding a visit to Chemnitz. 
Authors are grateful to Shashwati Roy Majumder for helpful suggestions regarding the simulation.
Some of the simulations have been run on IBM P690 at the Mobile Computing Centre, 
Jadavpur University.

\begin{table}
\begin{center}
\begin{tabular}{|c|c|c|}
\hline
  & slip  &  no slip
\\
\hline
S2& $t_d$ for cross-over & $t_d$ for cross-over\\ \hline
0.04 & peel for all $t_d$ & peel for all $t_d$\\
0.05 & 75 ($\sim$ simultaneous peel and crack) &  peel for all $t_d$ \\
0.06 & 30-31 (peel$\rightarrow$crack) & 34-35 (peel$\rightarrow$crack) \\
0.07 &  crack for all $t_d$ &  crack for all $t_d$\\
\hline
$t_d$ & S2 for cross-over & S2 for cross-over \\
\hline
1 & 0.06-0.065 (peel-crack) & 0.065-0.07 (peel-crack)\\ 
\hline 
\end{tabular}
\caption{Cross-over regimes from peel to crack for different slip conditions on $L=20$ samples.
$t_d$ has been varied from 1 to 80.}\label{slip}
\end{center}
\end{table}

\newpage
\newpage
                                                                                
\newpage

\newpage
                                                                                                    
\newpage
                                                                                                    


\begin{thebibliography}{99}
                                                                                                                             
\bibitem{coffee}A. Groisman, E. Kaplan, Europhys. Lett. 25 (1994) 415
                                                                                                                             
\bibitem{shor}K.A. Shorlin,J.R. de Bruyn, Phys. Rev. E 61 (2000) 6950-6957
\bibitem{born}S. Bohn, J. Platkiewicz, B. Andreotti, M. Adda-Bedia, Y. Couder,
 71 (2005), 046215
\bibitem{pats}T.S. Komatsu, S. Sasa, J. Appl. Phys. 36 (1997) 391-395
                                                                                                    \bibitem{dm1}D. Mal, S. Sinha, S. Mitra, S. Tarafdar, Physica A
346 (2005) 110-115
                                                                                                                             
\bibitem{dm2}D. Mal, S. Sinha, S. Mitra, S. Tarafdar, Fractals 14 (2006)
283-288
                                                                                                    \bibitem{dm3}D. Mal, S. Sinha, T. Dutta, S. Mitra, S. Tarafdar, J .Phys. Soc.
Japan, 76 (2007) 014801(5 pages)
                                                                 
\bibitem{kit}S. Kitsunezaki, Phys. Rev. E,{\bf 60}, (1999) 6449\\
\bibitem{col1} H.Colina, L. Arcangelis, S. Roux, Phys. Rev. B, {\bf 48},
(1993), 3666\\
\bibitem{col2} H. Colina, S. Roux, Eur.Phys. J. E ,
 {\bf 1}, (2000), 189\\
                                                            
\bibitem{sadhu} S. Sadhukhan S; S.Roy Majumder , D. Mal, T. Dutta, S.Tarafdar, J. Phys. Condens.
Matter, {\bf 19}, (2007) 356206\\
\bibitem{wein} R. Weinberger, J. Struct. Geol., {\bf 21},(1999)  379\\
\bibitem{epje}S. Sinha, T. Dutta and S. Tarafdar, Eur. Phys. J. E {\bf 25}, (2008) 267 .

\end{thebibliography}
\end{document}